\documentclass[fleqn,usenatbib,]{mnras}

\usepackage{newtxtext,newtxmath}

\usepackage[T1]{fontenc}

\DeclareRobustCommand{\VAN}[3]{#2}
\let\VANthebibliography\thebibliography
\def\thebibliography{\DeclareRobustCommand{\VAN}[3]{##3}\VANthebibliography}
\usepackage[utf8]{inputenc}
\usepackage{amsmath}

\usepackage{siunitx}
\usepackage{graphicx}
\usepackage{float}
\usepackage{subcaption}

\newcommand{\insight}{\textit{Insight}-HXMT}
\newcommand{\swj}{Swift J0243.6+6124}

\title[\insight{}/ME dead time]{Revisiting the dead time effects of \insight{}/ME on timing analysis}

\author[Y. Tuo et al.]{
Youli Tuo,$^{1}$
Xiaobo Li, $^{2}$
\thanks{Contact e-mail: \href{mailto:lixb@ihep.ac.cn}{lixb@ihep.ac.cn}}
Ying Tan, $^{2}$
Baiyang Wu, $^{2}$
Weichun Jiang, $^{2}$
Liming Song, $^{2}$
\newauthor Jinlu Qu, $^{2}$
Sudeep Gogate, $^{1}$
Shuang-Nan Zhang, $^{2}$
Andrea Santangelo, $^{1}$
\\
$^{1}$Institut f{\"u}r Astronomie und Astrophysik, Universit{\"a}t T{\"u}bingen, Sand 1, D-72076 T{\"u}bingen, Germany\\
$^{2}$Key Laboratory of Particle Astrophysics, Institute of High Energy Physics, Chinese Academy of Sciences, 19B Yuquan Road, Beijing 100049, China\\
}

\date{Accepted XXX. Received YYY; in original form ZZZ}

\pubyear{2015}

\begin{document}
\label{firstpage}
\pagerange{\pageref{firstpage}--\pageref{lastpage}}
\maketitle

\begin{abstract}
Dead time is a common instrumental effect of X-ray detectors which would alter the behavior of timing properties of astronomical signals, such as distorting the shape of power density spectra (PDS), affecting the root-mean-square of potential quasi-periodic oscillation signals, etc. We revisit the effects of the dead time of Medium Energy X-ray telescope (ME) onboard \insight{}, based on the simulation of electronic read-out mechanism that causes the dead time, and the real data. We investigate dead time effects on the pulse profile as well as the Quasi-Periodic Oscillation (QPO) signals. The dead time coefficient suggests a linear correlation with the observed count rate in each phase bin of the pulse profile according to the simulation of periodic signal as well as the real data observed on \swj{}. The Fourier-amplitude-difference (FAD) method could well recover the intrinsic shape of the observed PDS in the case that the PDS is from two identical detectors. We apply this technique on ME, by splitting the 9 FPGA modules into 2 groups. The results indicate that the FAD technique suits the case when two groups of detectors are not largely different; and the recovered PDS of Sco X-1 observed by ME slightly enhances the significance of the previously known QPO signal, meanwhile the root-mean-square of QPO is significantly improved. We provide the FAD correction tool implemented in HXMTDAS for users in the future to better analyze QPO signals.
\end{abstract}


\section{Introduction}
The Hard X-ray Modulation Telescope (HXMT), dubbed \textit{Insight}-HXMT, was successfully launched on June 15, 2017 \citep{Zhang_2020,li2020flight}. It is China’s first X-ray astronomy satellite devoted to broad band observations in the 1--250\,keV. It carries three main telescopes: High Energy X-ray telescope\,(HE) using an array of NaI(Tl)/CsI(Na) scintillation detectors \citep{liu2019high}, covering the 20--250\,keV energy band, Medium Energy X-ray telescope\,(ME) using an array of Silicon Positive-Intrinsic-Negative (Si-PIN) detectors in the 5--30\,keV band \citep{cao2019medium}, and Low Energy X-ray detector\,(LE) using an array of Swept Charge Device (SCD) detectors in the 1--15 keV band \citep{chen2020low}.

In general dead time refers to the period during which the detector is unable to respond to new incoming signals. Dead time is classified into two categories: paralyzable and non-paralyzable. In the paralyzable type, every incident event, even if undetected, contributes to the dead time, making this type cumulative. As a result, paralyzable dead time can, in theory, become infinitely long, as each new incident event extends the unresponsive period. On the other hand, non-paralyzable dead time occurs only when an event is detected. Unlike the paralyzable type, subsequent undetected events during this period do not prolong the total dead time. 

\citet{1995ApJ...449..930Z} gave the precise models for both paralyzable and non-paralyzable dead time. While the dead time was assumed to be constant for non-paralyzable model, that is not true in practice. This model has been used to help calculate the corrected Power Density Spectrum (PDS, the squared Fourier amplitude) and root mean square (RMS) in some X-ray detectors. For example, it was used to compute the dead-time affected Poisson noise power and RMS for AstroSat/LAXPC data \citep{agrawal2018spectral,sreehari2020astrosat}. For Rossi X-Ray Timing Explorer (RXTE), based on this model, some extra empirical formulas were used to estimate dead time affected PDS considering different dead time produced by different types of events \citep{jernigan2000discovery,jahoda2006calibration}.
Apart from the dead time model, NuSTAR, with its two independent and identical detectors, has been analyzed using a method proposed by \citet{bachetti2015no}. This method employs the cospectrum, which is the real part of the cross PDS measured by two identical detectors, to obtain an accurate representation of the white-noise-subtracted PDS. Furthermore, \citet{bachetti2018no} suggested the Fourier amplitude differences (FAD) method to normalize dead-time affected cospectra and PDS. Their approach uses the difference of the Fourier amplitudes from two independent and identical detectors to characterize and filter out the effect of dead time for \textit{NuSTAR}. 
We revisit and investigate the dead time effects of the \insight{}/ME telescope. The paper is organized as follows. Section \ref{sec:read-out} introduces the detailed read-out mechanism responsible for the dead time in \insight{}/ME. In Section \ref{sec:eventsim}, we describe the simulation algorithm and methodologies that mirror the read-out mechanism, as well as the timing products prepared for further simulation. Section \ref{sec:FAD} details the adaptation and implementation of the FAD correction method specifically optimized for \insight{}/ME. In Section \ref{sec:sim-results}, we present and evaluate the simulation results. Section \ref{sec:hxmt-fad} discuss the validation of the specialized FAD method when applied to \insight{}/ME. Finally, we summarize the paper in Section \ref{sec:sum}.

\section{Dead time in \insight{}/ME}\label{me_dt}
\subsection{electronic read-out mechanism}\label{sec:read-out}

The ME electronics utilizes Application Specific Integrated Circuit (ASIC) chips to collect, amplify, digitize, and transmit the output signals of the detectors. There are 54 ASIC chips, each of which contains 32 channels, to meet the requirements of the read-out system of all the Si-PIN detectors. Every six ASICs in one detector unit are configured and controlled by one Field Programmable Gate Array (FPGA).

\begin{figure}
    \centering
    \includegraphics[width=\linewidth]{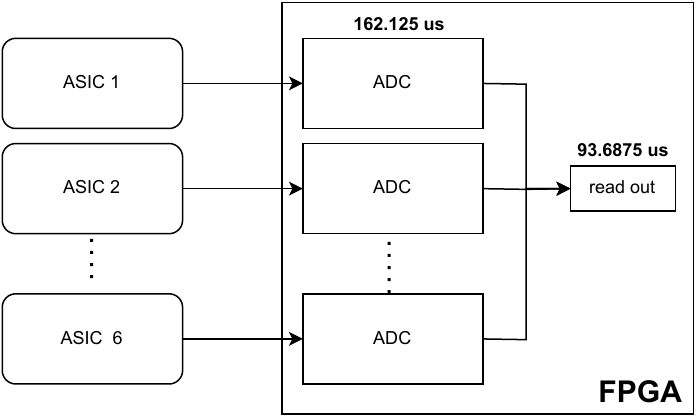}
    \caption{The schematic processes of \insight{}/ME read-out system.}
    \label{fig:read-out}
\end{figure}

\begin{figure*}
    \centering
    \includegraphics[width=0.8\textwidth]{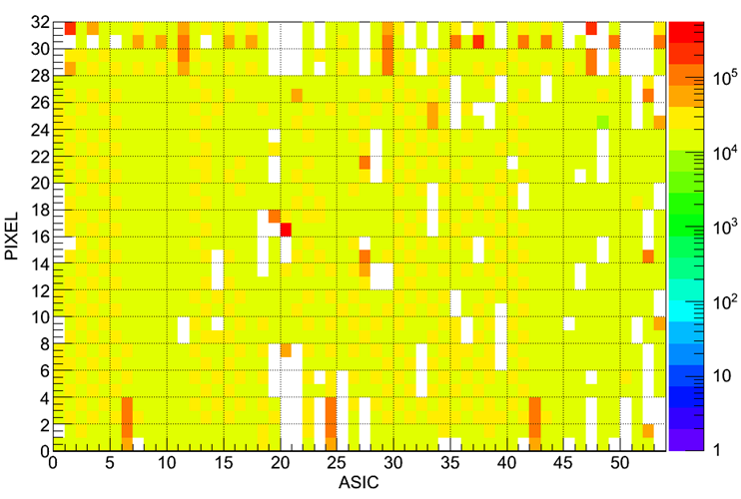}
    \caption{Pixels and their working situations of 9 FPGAs, where each FPGA contains 6 ASICs, and each ASIC includes 32 pixels. The pixels shown in white indicate 'bad pixels' in ME telescope, while the pixels highlighted in red, which exhibit a high count rate, are those attached with the calibration sources.}
    \label{dt:ME_fpga_pixel}
\end{figure*}

The dead time during signal acquisition is constituted by both the Analog-to-Digital (AD) conversion time and the read-out time (as schematised in Figure \ref{fig:read-out}). Specifically, the AD time corresponds to 2594 cycles of crystal oscillation, and the read-out time equates to 1499 cycles of crystal oscillation. With a 16\,MHz crystal oscillating at its nominal period of \SI{62.5}{\nano\second} at room temperature, the AD time and read-out time can be quantified as \SI{162.125}{\micro\second} and \SI{93.6875}{\micro\second}, respectively \citep{cao2019medium, li2022quasi}. The total read-out time for $n$ triggered ASICs, and thus the theoretical dead time value for the ME detector in terms of electronics, can be expressed by the following equation:
\begin{equation}\label{eq:medeadtime}
    {t}_{{\rm d}}={t}_{\rm{adc}}+{n} \times {t}_{{\rm rd}},
\end{equation}
where ${t}_{{\rm d}}$ is the total dead time, $t_{\mathrm{adc}}$ is AD conversion time, and $n$ is the number of ASICs that are triggered. If $n$ ASICs out of six ASICs sharing one FPGA are triggered during the same FPGA sampling period (\SI{62.5}{\nano\second}), they are considered to be triggered simultaneously. The typical value of $t_{\rm adc}$ is \SI{162.1}{\micro\second} and that of $t_{\rm rd}$ is \SI{93.7}{\micro\second}. Since the in-orbit count rate is quite low and only one ASIC is triggered most of the time, ${t}_{{\rm d}}$ of ME is about \SI{256}{\micro\second} in most cases \citep{cao2019medium}. The dead time in each FPGA is independent. Thus, the dead time of one ME FPGA is considered to be close to non-paralyzable. In Figure \ref{dt:ME_fpga_pixel} the real triggered counts for each pixel of each ASIC among all onboard ASICs are displayed in the 2-D diagram.

\subsection{simulation framework of the ME dead time}\label{sec:eventsim}
As described above, the value of the dead time in the case of ME is not a static value, but rather determined dynamically for each photon through a specialized sampling process. In this section we introduce the simulation processes that emulate the dead time effect of ME, as well as the timing products of both periodic and quasi-periodic signals. 

\begin{figure*}
    \centering
    \includegraphics[width=\linewidth]{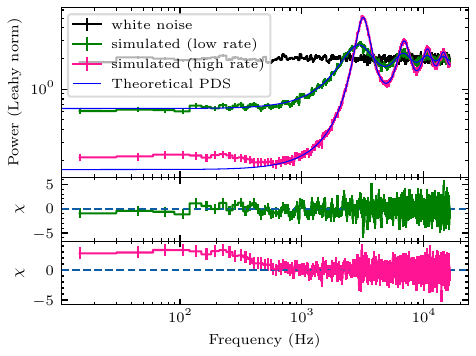}
    \caption{The periodograms of the white noise signal in different cases are illustrated with various color-coded lines. Black lines represent the PDS of pure white noise without any effects from dead time. The green lines depict the simulated PDS under a low count rate condition (3000 cnts/s), as affected by the ME dead time mechanism. The pink lines illustrate the same signal but under a high count rate condition (10000 cnts/s). The blue line is used to theoretically describe the PDS as distorted by a fixed dead time value of \SI{256}{\micro\second} in a non-paralyzable system. The second and the third panels are the residuals between the simulated PDS and the values predicted by models.}\label{fig:sim_pds}
\end{figure*}

We start with the straightforward case of simulating a white noise signal affected by dead time as shown in Figure \ref{fig:sim_pds}. This process begins by defining a count rate, from which a time series of photon arrival times is generated following a Poisson process. We create photon arrival times based on a Poisson process with an exponential distribution between arrival times of two photons. The arrival times are sampled according to the given rate, and an ASIC value is randomly assigned to each photon. These ASIC values are uniformly sampled from the set  \{1,2,3,4,5,6\}, reflecting the equal probability that a photon will incident on each ASIC. Subsequently, the dead time is computed based on the number of ASIC values ($n$) detected within a 62.5-nanosecond time window around each photon's arrival. Specifically, the dead time for a given photon is determined according to Equation \ref{eq:medeadtime}. 

To accelerate the computational process, we divide the time series into segments, each containing 1000 photons. We then perform parallel dead time filtering on each segment and carefully examine the photons at the boundaries of each segment to avoid potential boundary problem, which refers to the case of photons at the boundary of each segment not being correctly filtered. This dynamic calculation of dead time takes into account the interactions and dependencies among consecutive photons. It offers a more accurate representation of the dead time effect compared to the standard non-paralyzable case, enhancing the understanding of the underlying process, especially for high count rates, where the value of the dead time can deviate from \SI{256}{\micro\second}.

Subsequently, we investigate the impact of dead time on a periodic signal. We construct a pulse profile consisting of random numbers of Gaussian components, utilizing the \texttt{PYTHON} function \texttt{draw\_random\_pulse} implemented in package \texttt{tatpulsar.simulation.profile\_sim}\footnote{The repository containing the relevant code and functions can be found hosted at https://github.com/tuoyl/tat-pulsar.}. We simulated the arrival times of a profile using the Inverse Cumulative Distribution Function (CDF) method (see e.g. \cite{bachetti2024fourier}). Given the simulated phase, reference time, and the signal's period, we defined a lightcurve by repeating the pulse profile across the observation period. The light curve is then normalized to act as a probability density function. We computed the cumulative distribution function (CDF) by numerically integrating the PDF. Random values uniformly distributed between 0 and 1 were generated and mapped to arrival times by inverting the CDF, ensuring that the simulated arrival times accurately reflect the periodic structure of the pulse profile.

We extract a series of photon arrival times that echo a periodic signal and apply a filter based on the dead time mechanism. This process yields two distinct simulated pulse profiles: one affected by the dead time effects and another free from these effects.

To substantiate the reliability of both the simulation process and the dead time correction methodologies in terms of pulsation analysis, we compare the simulated pulse profile with real data. We require two profiles that are generated from the real data: one affected by dead time effects and another is dead-time unaffected. The HXMT data analysis procedures provide the methods of dead time corrections on the light curves. For each time bin in the light curves, the dead time ratio is computed by counting the triggered ASICs and applying Equation \ref{eq:medeadtime}. We generate light curves for each FPGA focusing on the source Swift J0243.6+612, maintaining a time resolution of one second. Following this, we compute the dead time ratio on each light curve and proceed to epoch-fold the light curve (function \texttt{tatpulsar.pulse.fold.fold\_lightcurve()} implemented in \texttt{tatpulsar}) into the pulse profile, using the timing parameters indicated on the GBM accreting pulsar histories\footnote{the monitored period of \swj{} could be found in https://gammaray.nsstc.nasa.gov/gbm/science/pulsars.html}.


\begin{figure}
    \centering
    \begin{subfigure}[htbp]{\linewidth}
    \includegraphics{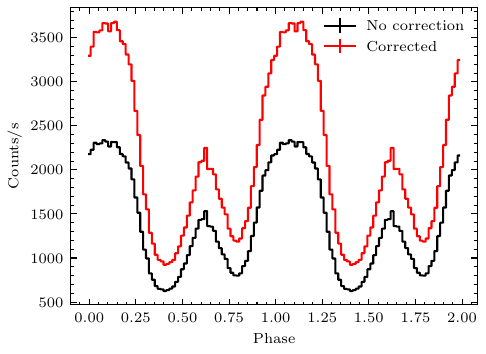}
    \caption{The pulse profile of the X-ray binary, \swj{}, as observed by HXMT. The profile, with a total exposure time of approximately 5100 seconds, was observed around the peak of this source's 2020 outburst. The black line represents the profile captured by HXMT without any correction for dead time, while the red line illustrates the profile post dead time correction.}
    \label{fig:0243profile}
    \end{subfigure}
    \hfill
    \begin{subfigure}[htbp]{\linewidth}
    \includegraphics{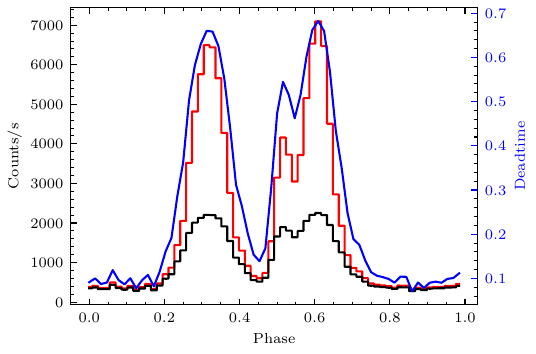}
    \caption{The simulated pulse profile}\label{fig:sim_profile}\label{fig:sim-profile-deadtime}
    \end{subfigure}    \begin{subfigure}[htbp]{\linewidth}
    \includegraphics{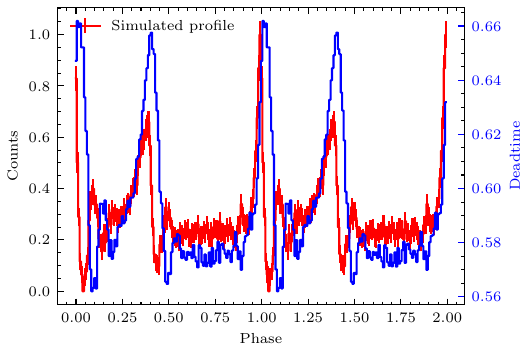}
        \caption{The simulated pulse profile of the Crab pulsar, with a dead time effect of \SI{2.5}{\milli\second}, shows that the peak of the dead time curve occurs in the delayed phase following the peak of the profile. The oscillatory pattern is similar to those observed in NuSTAR \citep{madsen2015broadband}, attributed to the cross-talk caused by the significant dead time.}\label{fig:profile_crosstalk}
	\end{subfigure}
    \caption{Pulse profiles.}
\end{figure}

\begin{figure}
    \centering
    \includegraphics{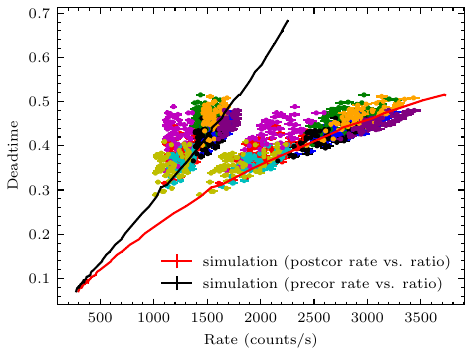}
    \caption{The correlation between the count rate and the dead time coefficient on the pulse profile. The black line represents the simulated correlation between the pulse profile count rate post dead time suppression and the corresponding dead time coefficient. The red line is the simulated correlation between the pulse profile count rate before dead time suppression and the corresponding dead time coefficient. The colored dots denote actual measurements taken from the observed pulse profile of the accreting pulsar \swj{}.}
    \label{fig:ratecor-profile}
\end{figure}

\subsection{The FAD correction}\label{sec:FAD}
We then investigated FAD (Fourier Amplitude Difference) method with ME observations. The FAD algorithm also has been implemented in Stingray \cite{huppenkothen2019stingray}. The FAD method requires two or more synchronous light curves with identical and independent dead time mechanisms. We employed the established FAD method, adapted specifically for the intricacies of the HXMT detector design. This method necessitates the utilization of two light curves that are observed by two identical detectors, both being affected by dead time. The Fourier transform of two light curves are denoted as $\mathcal{F}_1$, and $\mathcal{F}_2$. Given that the light curve is the convolution of the dead time filter $d(t)$ with the physical originated signal $s(t)$ plus noise $n(t)$, it can be expressed as $f(t) = [(s(t) + n(t)] * d(t)$. According to the convolution theorem, the Fourier transform of the light curve is:
\begin{equation}
    \mathcal{F}(\nu) = [(\mathcal{S}(\nu) + \mathcal{N}(\nu)]\cdot\mathcal{D}(\nu),
\end{equation}
where $\mathcal{S}(\nu)$, $\mathcal{N}(\nu)$, and $\mathcal{D}(\nu)$ are the Fourier transform of the signal, noise, and dead time filter, respectively. For two identical detectors, the physical signal $\mathcal{S}(\nu)$ is identical for both detectors, it cancels out in the difference of the Fourier amplitudes of their light curves:
\begin{align}
    \mathcal{F}_1 - \mathcal{F}_2 &=(\mathcal{S}_1 + \mathcal{N}_1)\cdot\mathcal{D}_1 - (\mathcal{S}_2 + \mathcal{N}_2)\cdot\mathcal{D}_2\\
    &= (\mathcal{N}_1 - \mathcal{N}_2)\cdot \mathcal{D} \quad (\text{if } \mathcal{D}_1 = \mathcal{D}_2 = \mathcal{D}).
\end{align}
The Fourier difference then only contains the dead time effects on the noise component, as the dead time filter $\mathcal{D}(\nu)$ is a multiplicative factor in the Fourier transform. This modulation of the dead time effects on the noise implies that the single-detector Fourier amplitudes are proportional, on average, to the difference of the Fourier amplitudes in different realizations, with a constant factor of $1/\sqrt{2}$ \citep{bachetti2018no}. Thus, the technique described in \cite{bachetti2018no} aims to correct the distortion of the power spectrum and/or cospectrum modulated by dead time effects. This is based on the fact that the FAD imprints the same dead time modulation on the noise only, and the power spectrum is proportional to the FAD\footnote{We note that the FAD is valid only when $\mathcal{D}_1=\mathcal{D}_2$. The averaged Fourier transform of the dead time filter over a large number of realizations meets the requirement of a constant factor of $1/\sqrt{2}$ between the FAD and the Fourier amplitude. However, when the number of realizations to average is less than $\sim$50, it is not a good approximation that $\mathcal{D}_1 = \mathcal{D}_2$, i.e., the proportionality does not hold (see related code in \url{https://doi.org/10.5281/zenodo.12691527}).}.


For the NuSTAR observatory, it employs two identical focal plane detectors, FPMA and FPMB \citep{harrison2013nuclear}. In this configuration, the Fourier amplitude differences of the PDS, observed by the two detectors, effectively normalize the periodogram affected by dead time \citep{bachetti2018no}. Meanwhile, for the \insight{}/ME configuration, there are nine FPGA modules, each equipped with an independent dead time processing unit as described in section \ref{sec:read-out}. To implement the FAD method on ME, and to make full use of the data observed by all FPGA modules, we divide the data from nine FPGA detectors into two groups. These groups were treated as two identical detectors that exhibit identical dead time effects. We iterate all potential grouping strategies of the nine FPGAs, eventually selecting a detector grouping strategy that equalizes the average count rate of the two groups as closely as possible, thereby making the two groups of detectors approximately identical as much as possible. In this way we apply the FAD method to the ME observations of Sco X-1, in which there is a QPO feature at around 800\,Hz \citep{jia2020insight}. We calculate the PDS in the time resolution of 1/4096 seconds, and the photons are extracted in the energy range of 8--30\,keV. The information of those observations is listed in Table 1 in \cite{jia2020insight}. 

The FAD technique recovers the significance of a QPO signal that might otherwise be suppressed by dead-time distortion. To further assess the performance, we conduct simulations in which a single QPO is detected in the PDS across various central frequencies. We fix the RMS of each QPO at 0.1 and sample the central frequency of the QPO from 10\,Hz to 850\,Hz, and sample the incident count rate between 1500\,cnts/s to 3500\,cnts/s. The QPO simulation technique has been implemented in \texttt{Stingray} \citep{huppenkothen2019stingray}, where the light curve is simulated by inverse Fourier transfer on the model of PDS with a QPO signal. After that, the desired RMS is achieved by scaling the simulated light curve \citep{timmer1995generating}. Then, we computed the significance of each simulated QPO to evaluate the FAD method's efficacy in restoring the QPO signal's significance. The significance of the detected QPO is quantified by dividing the best-fit normalization of the Lorentzian function by the corresponding error in normalization. 

\begin{figure*}
    \centering
    \includegraphics[width=\linewidth]{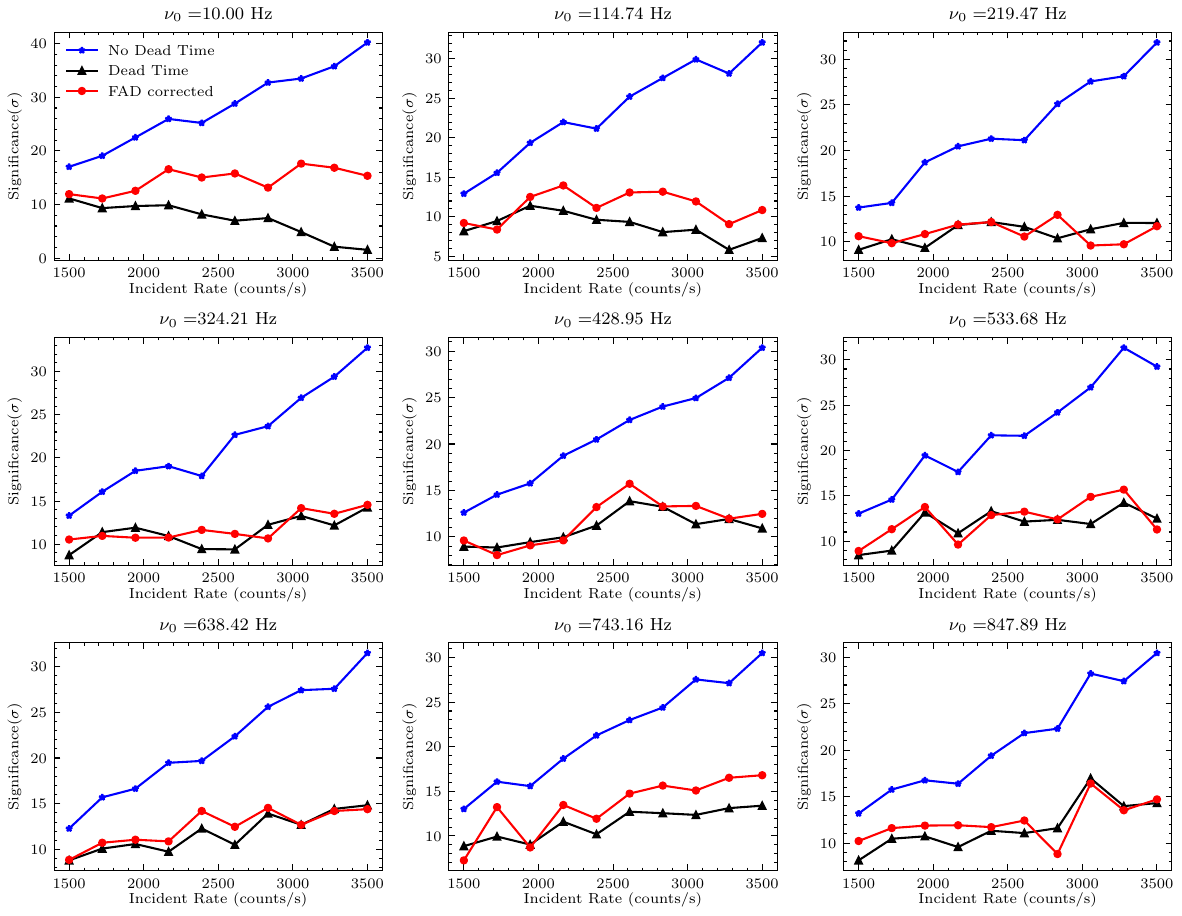}
    \caption{The simulated QPO significance across various conditions, as shown in each panel. Blue asterisk represents the simulated QPOs unaffected by dead time, while black triangles indicate the QPO significance suppressed due to simulated dead time effects from the HXMT/ME instrument. Red circles show the significance recovered using the FAD method when two FPGAs are taken into account. Each panel shows the variation in significance with respect to the incident count rate of simulated light curves. Every panel represents a different case, characterized by a different central frequency of the simulated QPO signal. Throughout the simulation, other QPO properties (such as RMS and FWHM) are fixed.}
    \label{fig:sim-qpo-significance}
\end{figure*}

\begin{figure*}
    \centering
    \includegraphics[width=\linewidth]{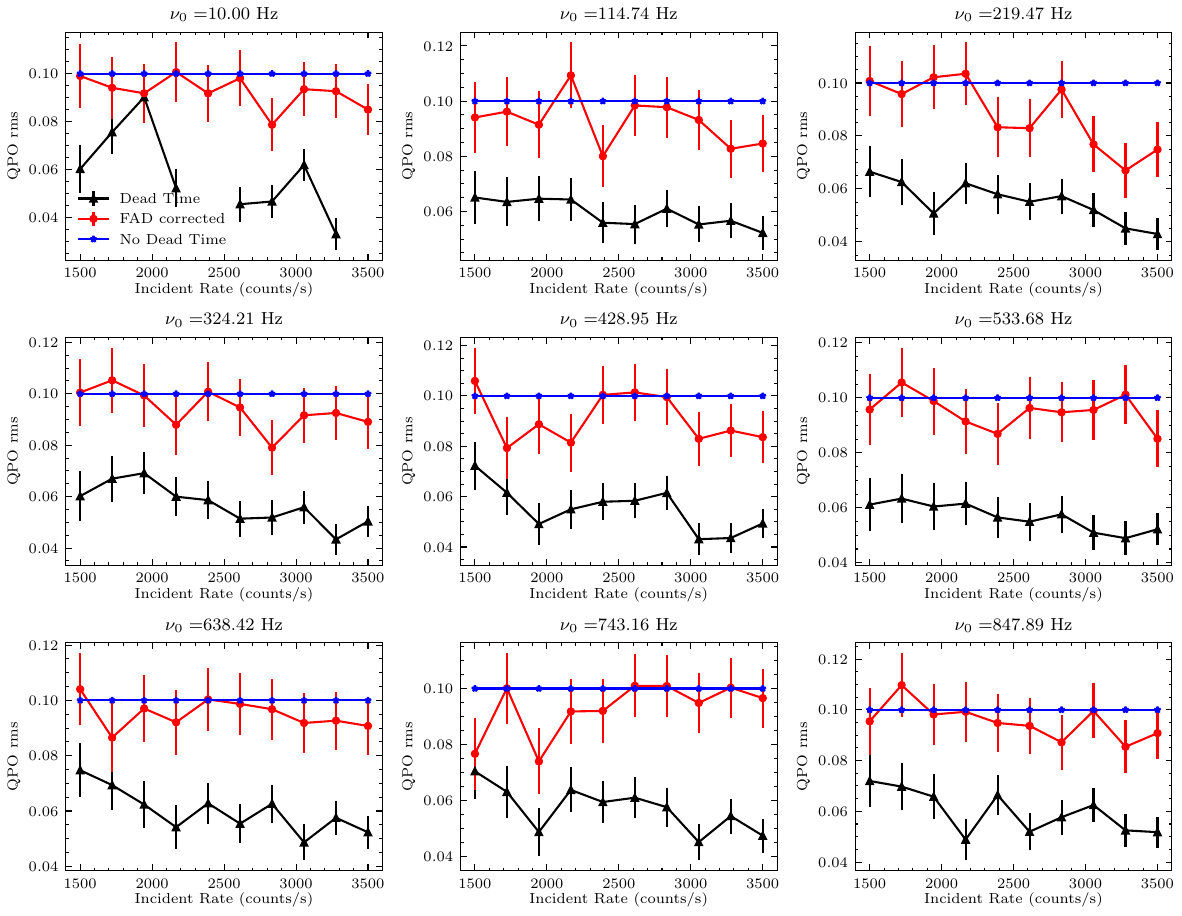}
    \caption{The simulated QPO RMS across various conditions, as shown in each panel. Blue asterisk represents the simulated QPOs unaffected by dead time, while black triangles indicate the RMS of QPO suppressed due to simulated dead time effects from the HXMT/ME instrument. Red circles show the QPO RMS recovered using the FAD method. Each panel shows the variation in RMS with respect to the incident count rate of simulated light curves. Every panel represents a different case, characterized by a different central frequency of the simulated QPO signal. Throughout the simulation, other QPO properties (such as RMS and FWHM) are fixed.}
    \label{fig:sim-qpo-rms}
\end{figure*}


\begin{figure}
    \centering
    \includegraphics{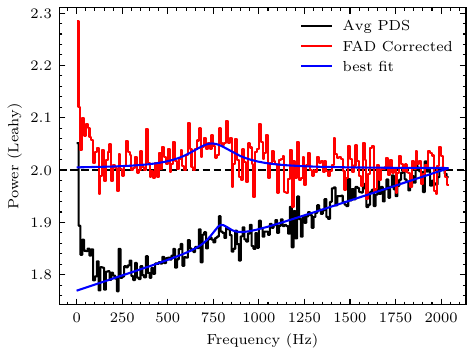}
    \caption{A typical periodogram of Sco X-1 observed by HXMT/ME. The black lines represent the PDS derived directly from the lightcurve from all 9 FPGAs of HXMT/ME. The red lines are the averaged FAD spectra, acquired through the calculation of the FAD utilizing two distinct PDS generated from two separate sets of FPGA data (further details are described in the context). The blue lines are the best fitting QPO signal model, applicable to both the FAD corrected PDS and the original PDS as detected by HXMT/ME. The black dashed line is the expected constant level set at a value of 2 as per the Leahy normalization method.}
    \label{fig:FAD-QPO}
\end{figure}

\section{results and discussion}\label{sec:discussion}
\subsection{simulation results}\label{sec:sim-results}
We first explore the simulation of the dead time mechanism on white noise, employing an instrumental effect-based simulation method. Specifically, this method simulates the behavior of individual photons in line with the read-out mechanism of the HXMT/ME instruments. This in-depth approach reveals that at high count rates, the dead time distortion on white noise may differ from traditional paralyzable or non-paralyzable cases. In Figure \ref{fig:sim_pds}, the black PDS is the white noise without any dead time effects, while the green and pink lines are the dead-time distorted PDS in relatively low count rate (3000 counts/s) and in high count rate (10000 counts/s), respectively. It appears that at high count rate, the dead time model \citep{1995ApJ...449..930Z} is limited to fit the component appearing in low-frequency range on PDS due to the \insight{}/ME read-out mechanism.  Nevertheless, given that the simulation operates at the FPGA unit level, the actual effective area cannot facilitate as high a count rate for each FPGA module as depicted in Figure \ref{fig:sim_pds}. Even in the instance of an ultra-luminous x-ray source such as \swj{}, the maximum flux recorded from a single FPGA module—approximately 3500 counts/s— occurred at the peak of its 2020 outburst. This count rate agrees closely with the low flux scenarios simulated in our simulation.

Subsequently, we aim at evaluating the validity of this correction approach through simulation. Specifically, we generate an event data set with a periodic signal, simulate the effects of dead time, and then compare the pulse profile with the ones folded by dead time corrected light curves. In Figure \ref{fig:0243profile}, the black profile illustrates the pulse profile of \swj{} affected by dead time, while the red profile presents the unaffected pulse profile. The pulse profile corrected for dead time has been folded by the corresponding light curve at a time resolution of 1 second. This light curve correction relies on the dead time coefficient determined by the number of triggered ASICs, as described in section \ref{sec:read-out}. This involves transitioning from a time domain analysis to the phase domain, utilizing the pulse epoch folding algorithm on light curves. The simulation progression of dead time is not only intuitive but also agrees with the instrumental characteristics. 

As shown in Figure \ref{fig:sim-profile-deadtime}, both the red and black pulse profiles represent simulated results. However, the dead time effect has been incorporated into the black pulse profile at the event level based on the simulation processes mentioned in section \ref{sec:eventsim}. We then compare the dead time effects across two distinct scenarios: one derived from simulation and the other from real data of \swj{}. The dead time ratio, $d$ is computed by $d=(r'-r)/r'$, where $r'$ is the count rate unaffected by the dead time, and $r$ denotes the count rate with the dead time effect. As shown in Figure \ref{fig:ratecor-profile}, the black line represents the simulated dead time coefficient corresponding to the pulse profile's count rate in each bin without the dead time correction. In contrast, the red line indicates the correlation between the dead time coefficient and the count rate unaffected by dead time in each bin. The colored dots represent the respective values measured for each HXMT/ME FPGA. The agreement between the simulated trend and the actual data measurements is notably consistent. This indicates that the simulated dead time ratio maintains a positive correlation with the observed count rate of the pulse profile in each phase bin. While the distribution from the actual data has the same correlation, and agrees to the simulated correlation, it does exhibit systematic uncertainties when determining the dead time effects based on the observed count rate for each FPGA. This correlation could be used to correct the dead time effect for the given count rate of a light curve or a pulse profile. The deviation of certain FPGA from the simulation line can be attributed to the simulation not accounting for the presence of bad pixels. We thus model the cases where photons, triggered by these specific bad pixels, are discarded before the dead-time filtering process. The trend resulting from this case, when combined with instances that do not involve pixel discarding, agrees more closely with the observed data, adequately addressing the systematic uncertainties. 

We also note that for short-period pulsations, the dead time triggered in one phase bin can impact the counts of other phase bins (i.e. cross-talk). This effect is particularly significant for short-period pulsations (e.g. the Crab pulsar) observed with NuSTAR \citep{madsen2015broadband}, which showcases a distorted livetime distribution along the pulsar phase. As illustrated in Figure \ref{fig:profile_crosstalk}, we simulated a dead time of \SI{2.5}{\milli\second} on the profile of the Crab pulsar with a period of \SI{33}{\milli\second}. The simulated profile of the Crab pulsar exhibited significant distortion in the observed pulse profile compared to the real one after accounting for the dead time response. And a delayed dead time of \SI{1.29}{\milli\second} between the livetime distribution and the observed profile is also significant. However, when simulating the case for HXMT/ME with a dead time of \SI{256}{\micro\second}, the cross-talk effect of the dead time is neglectable, with a delay of \SI{29.9}{\micro\second}. Consequently, the observed Crab profile shows no distortion, which is consistent with the on-board Crab observations \citep{tuo2022orbit}, unlike the distortions seen in NuSTAR. Regardless, this demonstrates the reliability of both the dead time effect simulation process and the correction method for light curves implemented in HXMTDAS (HXMT data analysis software).

\subsection{FAD results on HXMT}\label{sec:hxmt-fad}
Investigating the dead time effect on timing analysis from the perspective of the QPO signal is important as well. Utilizing the specific FAD method for HXMT/ME, as described in section \ref{sec:FAD}, we assess its validity from both simulated and real data perspectives observed on Sco X-1.

As shown in Figure \ref{fig:sim-qpo-significance}, the significance of the simulated QPO signal in three different scenarios is computed, the blue asterisks are significance of dead-time unaffected QPO signals, the black triangles are the significance computed in dead-time distorted PDS, and the red circles are the significance corrected by FAD methods. Each panel shows the significance variation with respect to the incident count rate of the lightcurve. The FAD method effectively recovers QPO signals in different central frequencies for \insight{} in low frequency case. With the increase of the central frequency, the significance of the QPO could be slightly recovered in specific count rate. 

We also evaluated the performance of the FAD method in restoring the measured RMS of QPO signals. As shown in Figure \ref{fig:sim-qpo-rms}, the FAD method significantly restored the QPO RMS, suppressed by the dead time effect, to levels close to the simulated values without the dead time effect. Overall, while the FAD method only slightly restored the significance of QPO signals, it managed to fully restore the measured RMS of QPO to the real value compared to the dead-time distorted PDS.

While there remains some loss in significance due to the dead time effects, this discrepancy can be estimated using other simulation-based techniques, such as the Simulation-Based Inferences (SBI) as introduced by \cite{huppenkothen2022accurate}. This simulation indicates good performance in restoring the QPO significance and we apply this method on real HXMT observations. 
In Figure \ref{fig:FAD-QPO}, the red lines are the FAD corrected PDS of Sco X-1 observation, where the significance for the QPO centered at $742.58\pm 24.58$\,Hz is $5.86\,\sigma$; while in \cite{jia2020insight}, the significance for this observation calculated in the periodogram with dead time distortion is $5.2 \pm 0.6\,\sigma$. Two observations from HXMT/ME detected kHz QPOs, as listed in Table \ref{table:significance-qpos}. These observations suggest that the FAD method not only effectively restores the significance of QPO signals in simulations but also agrees well with real data.

\begin{table*}
 \centering
    \begin{tabular}{ccccc}
        \hline
        obsid & FAD corrected & dead time distorted & FAD corrected & dead time suppressed \\
              & significance  & significance        & RMS           & RMS \\
        \hline
        P010132801001 & $5.86 \sigma$ & $5.2\sigma$ & $0.50\pm0.04$ & $0.05\pm0.04$ \\
        \hline
        P010132801002 & $5.28\sigma$ &  $4.2\sigma$ & $0.40\pm0.04$ & $0.05 \pm 0.04$\\
        \hline
    \end{tabular}
    \caption{The significance for FAD corrected QPO signal and the QPO significance calculated in dead time distorted periodogram.}
    \label{table:significance-qpos}
\end{table*}

\section{summary}\label{sec:sum}
We conducted a series of simulations on pulse profiles and QPO signals to explore the impacts of the dead time effect from HXMT/ME. The dead time simulation, based on the electronic read-out mechanism, accurately reflects the real data of observations with periodic signals. Moreover, the FAD method, specifically optimized for HXMT/ME, effectively restores QPO RMS to values close to the real ones and can also restore the QPO significance in certain cases.. To benefit the HXMT community, these simulation tools and the FAD method for HXMT/ME QPO detection have been integrated into HXMTDAS \footnote{http://hxmten.ihep.ac.cn/SoftDoc.jhtml}, and also pave the way for enhancing the accuracy of timing analysis for payloads onboard the future mission like e\textit{XTP} \citep{zhang2017extp}.

\section*{Acknowledgements}

The authors would like to express our sincere gratitude to the referee for their insightful comments and constructive suggestions. The thorough review and valuable feedback have significantly contributed to the improvement of our work. The authors thank supports from the National Natural Science Foundation of China under Grants 12273043, U2038109. This work used data from the \textit{Insight}-HXMT mission, a project funded by the Strategic Priority Research Program on Space Science of the Chinese Academy of Sciences and the China National Space Administration (CNSA). We gratefully acknowledge the support from the National Program on Key Research and Development Project (Grant No.2021YFA0718500) from the Minister of Science and Technology of China (MOST). 

\section*{data availability}
The data underlying this article are available in \insight{} achieve database, at \url{http://archive.hxmt.cn/proposal}. Pulse profile and PDS data presented in the various plots of the manuscript are available in Zenodo repository at \url{https://doi.org/10.5281/zenodo.12691527}.

\bibliographystyle{mnras}
\bibliography{ref}

\end{document}